%% file: main.tex
\documentclass[conference]{IEEEtran}
\usepackage{cite}
\usepackage{amsmath,amssymb,amsfonts}
\usepackage{algorithmic}
\usepackage{graphicx}
\usepackage{textcomp}
\usepackage{xcolor}
\usepackage{balance}
\usepackage{xspace}
\usepackage{pifont}
\usepackage{subcaption}
\usepackage{booktabs}
\usepackage{colortbl}
\usepackage{multirow}
\usepackage{url}
\usepackage{array,ragged2e}
\usepackage{booktabs}
\usepackage{enumitem}
\usepackage{tablefootnote}

\def\BibTeX{{\rm B\kern-.05em{\sc i\kern-.025em b}\kern-.08em
    T\kern-.1667em\lower.7ex\hbox{E}\kern-.125emX}}

\newcommand{\ie}{\emph{i.e.,}\xspace}
\newcommand{\eg}{\emph{e.g.,}\xspace}
\newcommand{\etc}{etc.\xspace}

\begin{document}

\title{Using Quality Attribute Scenarios for ML Model Test Case Generation}

\author{\IEEEauthorblockN{Rachel Brower-Sinning, Grace A. Lewis, Sebasti\'{a}n Echeverr\'{i}a, Ipek Ozkaya}
\IEEEauthorblockA{\textit{Carnegie Mellon Software Engineering Institute}\\
Pittsburgh, PA USA \\
\{rbrowersinning, glewis, secheverria, ozkaya\}@sei.cmu.edu}
}

\maketitle

\begin{abstract}
Testing of machine learning (ML) models is a known challenge identified by researchers and practitioners alike. Unfortunately, current practice for ML model testing prioritizes testing for model performance, while often neglecting the requirements and constraints of the ML-enabled system that integrates the model. This limited view of testing leads to failures during integration, deployment, and operations, contributing to the difficulties of moving models from development to production. This paper presents an approach based on quality attribute (QA) scenarios to elicit and define system- and model-relevant test cases for ML models. The QA-based approach described in this paper has been integrated into MLTE, a process and tool to support ML model test and evaluation. Feedback from users of MLTE highlights its effectiveness in testing beyond model performance and identifying failures early in the development process.
\end{abstract}


\input{introduction}
\input{quality-attributes}
\input{mapping}
\input{scenarios}
\input{benefits}
\input{application}

\input{related-work}
\input{summary}

\section*{Acknowledgments}
This material is based upon work funded and supported by the Department of Defense under Contract No. FA8702-15-D-0002 with Carnegie Mellon University for the operation of the Software Engineering Institute, a federally funded research and development center (DM24-0121).

\bibliographystyle{IEEEtran}
\bibliography{references.bib}

\end{document}

%% file: introduction.tex
\section{Introduction}\label{sec:introduction}

Testing of machine learning (ML) models has been identified by both researchers and practitioners as a challenge \cite{Nahar2023}. Current practice for ML model testing during ML model development prioritizes testing model properties, such as model performance (\eg accuracy), without adequate consideration of system requirements, such as throughput, resource consumption, or robustness, leading to failures in model integration, deployment, and operations \cite{Braiek2020}\cite{Golendukhina2022}. In many cases, model developers lack the skills or background to test beyond model performance, and in others model developers receive very little system context that informs design decisions\cite{Lewis2021}.

This paper is an experience report on the use of quality attribute (QA) scenarios for ML model test case generation, and the integration of the approach in a tool called MLTE that supports both system- and model-centric testing. Applications of this QA-driven ML model test case generation approach in practice confirm its benefit in developing and deploying models that better meet system needs and report less failures in production. Section \ref{sec:quality-attributes} provides a background on QA scenarios. Section \ref{sec:mapping} shows the mapping between QA scenarios and ML model test cases, with examples presented in Section \ref{sec:scenarios}, and benefits of the approach in Section \ref{sec:benefits}. Results of the application of the approach in practice are presented in Section \ref{sec:application}. Finally, related work is listed in Section \ref{sec:related-work}, and summary and next steps in Section \ref{sec:summary}.

%% file: quality-attributes.tex
\section{Quality Attribute Scenarios}\label{sec:quality-attributes}

Quality attribute (QA) scenarios have been used effectively as a way to specify a system's structural and behavioral requirements, referred to as quality attribute requirements \cite{Bass2021}. The goal of using scenarios is to ensure that QA requirements are described unambiguously and are testable. Software architects and developers use scenarios primarily to guide their design decisions, and as test cases to determine if QA requirements have been achieved. QA scenarios have six parts:

\begin{itemize}
\item Stimulus: A condition arriving at the system (\eg event, user operation, attack, request for modification, completion of a unit of development)
\item Source of Stimulus: Where the stimulus comes from (\eg internal/external user, internal/external system, sensor)
\item Environment: Set of circumstances in which the scenario takes place (\eg normal operation, overload condition, startup, development time)
\item Artifact: Target for the stimulus (\eg system, subsystem, component, data store, user interface, ML model)
\item Response: Activity that occurs as the result of the arrival of the stimulus (\eg process event, deny access, implement modification, test)
\item Response Measure: Measures used to determine that the responses enumerated for the scenario have been achieved (\eg latency, throughput, execution time, effort)
\end{itemize}

ML models are one category of architecture element in ML-enabled systems. Therefore, similar to other elements, QA requirements relevant to ML models need to be considered. These QA requirements include those related to the ML model as one of the elements of the system, as well as those relevant to how the ML model interacts with other elements in the system. We posit that QA scenarios need to be used as a way to specify these requirements for ML models, with the same goal of ensuring that ML model elements consider the system that they will operate in effectively, and that requirements are unambiguous and testable. In this case, model developers can use the QA scenarios for several purposes:
\begin{itemize}
\item communicate and negotiate with stakeholders to ensure that model requirements are clear, testable, and reproducible,
\item use as input to model design decisions,
\item reason about and guide trade-offs about model design, as well as how the model integrates with the system, and
\item build test cases, which is the focus of this paper.
\end{itemize}

In addition, adoption of QA scenarios brings architectural, system quality, and systems thinking to the ML model development process --- beyond just model performance --- and makes explicit how an ML model is expected to function and behave in production once it is part of a larger system. While this practice  is common in traditional software development, it is not in model development, which is typically done by data scientists not trained in these practices.

%% file: mapping.tex
\section{Mapping QA Scenarios to Test Cases}\label{sec:mapping}

Using QA scenarios to guide model development and test case generation follows a similar process as for other software components. After the model developer elicits, clarifies, and negotiates model requirements with relevant stakeholders (\eg product/system owner) using QA scenarios, the mapping of scenarios to test cases is done following Table \ref{tab:mapping}. The goal of the mapping is to develop test cases for each model requirement. Artifact is not included in the table because the artifact in this case will always be the model under test.

\begin{table} [hbt]
    \centering
    \caption{Mapping QA Scenarios to ML Model Test Cases}
    \begin{tabular}{|p{1.6cm}|p{1.6cm}|p{4.3cm}|}
    \hline
        \multicolumn{1}{|>{\centering\arraybackslash}p{1.6cm}|}{\textbf{QA Scenario}} &
        \multicolumn{1}{>{\centering\arraybackslash}p{1.6cm}|}{\textbf{ML Model Test Case}} &
        \multicolumn{1}{>{\centering\arraybackslash}p{4.3cm}|}{\textbf{Translation to ML Model Test Code}} \\ \hline
        Stimulus and Source of Stimulus & Data and Data Source & 
        \begin{itemize} [label=\textbullet, nolistsep, leftmargin=10pt,
            before*={\mbox{}\vspace{-\baselineskip}},after*={\mbox{}\vspace{-\baselineskip}}]
            \item Identification of dataset that meets specified characteristics (quality and quantity)
            \item Generation or transformation of data to meet scenario needs
        \end{itemize}  \\ \hline
        Response and Response Measure & Measurement and Condition per Response&
        \begin{itemize} [label=\textbullet, nolistsep, leftmargin=10pt,
            before*={\mbox{}\vspace{-\baselineskip}},after*={\mbox{}\vspace{-\baselineskip}}]
            \item Identification of test(s) to meet response measure
            \item Test setup 
            \item Definition of conditions for test to pass
        \end{itemize}    \\ \hline
        Environment & Context &
         \begin{itemize} [label=\textbullet, nolistsep, leftmargin=10pt,
            before*={\mbox{}\vspace{-\baselineskip}},after*={\mbox{}\vspace{-\baselineskip}}]
            \item Replication of any context conditions, such as operational environment, input failure, or higher data arrival rates
        \end{itemize} \\ \hline
    \end{tabular}
    \label{tab:mapping}
\end{table}

While many test cases will likely be data-driven and related to runtime model behavior, it is important to recognize that there will be scenarios that are more similar to traditional software scenarios, such as a maintainability scenario in which the stimulus is a request for change and the response is time to implement that change. In these cases, software QA scenarios can be used with their corresponding test cases.  

%% file: scenarios.tex
\section{Sample Scenarios and Test Cases}\label{sec:scenarios}

The examples in Table \ref{tab:scenarios} relate to a hypothetical system used by visitors to a botanical garden to identify flowers in the different sections of the botanical garden (\eg rose garden, succulent garden, \etc) and learn more about them. The system uses an ML model that was trained on the flower category dataset \cite{Nilsback2008}.  Each row in the table is a test case, built using the mapping from Table \ref{tab:mapping}. The \textit{Scenario Description} column is the elicited QA scenario for the model requirement. 

\begin{table*} [t!]
    \centering
    \caption{Sample QA Scenarios and Corresponding ML Model Test Cases}
    \begin{tabular}{|p{1.7cm}|p{4.4cm}|p{4.4cm}|p{4.2cm}|p{1.2cm}|}
    \hline
        \multicolumn{1}{|>{\centering\arraybackslash}p{1.7cm}|}{\textbf{Quality Attribute}} &
        \multicolumn{1}{>{\centering\arraybackslash}p{4.4cm}|}{\textbf{Scenario Description}} &
        \multicolumn{1}{>{\centering\arraybackslash}p{4.5cm}|}{\textbf{Data and Data Source}} &
        \multicolumn{1}{>{\centering\arraybackslash}p{4.2cm}|}{\textbf{Measurement and Condition}} &
        \multicolumn{1}{>{\centering\arraybackslash}p{1.2cm}|}{\textbf{Context}} 
        \\ \hline
        
        Fairness: Model Impartial to Photo Location & The model receives a picture taken at the garden and, regardless of the location in the garden, can identify the correct flowers at least 90\% of the time during normal operation. & Test data needs to include pictures of the flowers from the different sections, grouped by the section that the image was taken at. The quantity of the flower images should be representative of the garden section population they are taken from.  & The total accuracy of the model across each garden population should be higher or equal to 0.9. & Normal operation\\ \hline

        Robustness: Model Robust to Noise (Image Blur) & The model receives a picture taken at any garden section by a member of the public, and it is a bit blurry.  The model should still be able to successfully identify the flower at the same rate as non-blurry images during normal operation. & Test data needs to include blurred flower images.  Test blurred images will be created using ImageMagick. \cite{ImageMagick2024}. Three datasets will be generated, each with different amounts of blur: minimal blur, maximum blur, and in between minimal and maximum blur. & Blurry images are successfully identified at rates equal to that of non-blurred images. This will be measured using the Wilcoxon Rank-Sum test, with significance at p-value $<=0.05$. & Normal operation\\ \hline

        Robustness: Model Robust to Noise (Channel Loss) & The model receives a picture taken at a garden using a loaned device. These devices are known to sometimes lose a channel (\ie RGB channel). The model should still be able to successfully identify the flower at the same rate as full images, during normal operation. & Test data needs to include images with a missing channel. Test images will be generated by removing the R, G and B channels in the original test data using ImageMagick \cite{ImageMagick2024}, therefore producing three data sets. & Images with a missing channel are successfully identified at rates equal to that of original images. This will be measured using the Wilcoxon Rank-Sum test, with significance at p-value  $<=0.05$. & Normal operation\\ \hline

        Performance (on Operational Platform) & During normal operation, the model runs on the devices loaned out by the garden to visitors, without any errors due to unavailable resources. These are small, inexpensive devices with limited CPU power, as well as limited memory and disk space (512 MB and 128 GB, respectively). & The original test dataset can be used (\ie original test data set aside for testing during data split into training, validation, and test). & 
       \begin{enumerate} [nolistsep, leftmargin=10pt,
            before*={\mbox{}\vspace{-\baselineskip}},after*={\mbox{}\vspace{-\baselineskip}}]
       \item Executing the model on the loaned platform will not exceed maximum CPU usage of 30\% to ensure reasonable response time. CPU usage will be measured using \texttt{ps}.
       \item Memory usage at inference time will not exceed available memory of 512 MB. This will be measured using \texttt{pmap}.
       \item Disk usage will not exceed available disk space of 128 GB. This will be measured using by adding the size of each file in the path for the model code.
       \end{enumerate}
        & Normal operation\\ \hline

        Interpretability: Understanding Model Results & The application indicates the main features that were used to recognize the flower, as part of the educational experience. The app will display the image highlighting the most informative features in flower identification, in addition to the flower name, during normal operations. & The original test dataset can be used. & The model needs to return evidence, in this case a heat map implementing the \textit{Integrated Gradients} algorithm, showing the pixels that were most informative in the classification decision. This evidence should be returned with each inference. & Normal operation\\ \hline

    \end{tabular}
    \label{tab:scenarios}
\end{table*}

The code that implements these test cases for the corresponding scenarios is available at \url{https://github.com/mlte-team/mlte/tree/master/demo/scenarios}. It is structured using Jupyter Notebooks \cite{Jupyter2024} and uses the MLTE process and infrastructure\cite{Maffey2023}\cite{MLTE2023} for building, running, and collecting the results of the test cases (see Section \ref{sec:application}).

%% file: benefits.tex
\section{Additional Benefits of the Proposed Approach Beyond Test Case Generation}\label{sec:benefits}

As shown in Sections \ref{sec:mapping} and \ref{sec:scenarios}, QA scenarios can be used to build test cases for ML models. However, the negotiation and discussions that take place during QA elicitation have added benefits to ML model development, especially in organizations where model development and software development teams are independent, and members have a data science or software engineering background, respectively.

\textbf{To communicate and negotiate with stakeholders to ensure that model requirements are clear, testable, and reproducible.} In practice, it is quite common (and unfortunate) for requirements to be ambiguous, open to interpretation, or lacking detail to fully understand what needs to be developed. As a very simple example, when a requirement is stated as ``The system shall be secure'' there is not enough information to fully understand what this means from a development perspective. A scenario that clarifies that under normal operation the system will only grant access to registered users starts to more clearly translate to the need for an authentication system and a user registry, and the need for testing that only registered users are granted access to the system and non-registered users are not. This same practice needs to be used in ML model development, especially in light of QAs such as model fairness and explainability, which are less understood and have several relevant QA concerns. Model developers should iterate scenario definition with stakeholders to ensure that the necessary detail is specified. In some cases, that detail may come from the stakeholders and subject matter experts; but in other cases, the model developer may need to provide details of statistical tests or data characteristics that will be used to create valid and realistic test cases.

\textbf{As input to model design decisions.} As in traditional software, QA scenarios are important inputs to model design \cite{Bass2021}. QAs such as interpretability and integratabilty (ease of integration) will likely influence choice of model architecture, software packages, and languages. Security concerns may limit availability of data for training or during operation. Robustness and fairness may influence how data is processed prior to and during training. Clearly stating and prioritizing system and model QAs will enable model developers to make critical design choices early in the development process and avoid cost of rework. 

\textbf{To reason about trade-offs.} In software architecture it is well known that QAs will not be achieved in isolation \cite{Bass2021}, which also applies to ML model quality attributes.  For example, designing a model to be fair and accurate across different operating conditions may impact its ability to operate effectively on platforms with limited computing power to run the model. Using scenarios enables discussions about trade-offs at an earlier point in the software development cycle, and forces stakeholders to prioritize competing QAs. For example, stakeholders may decide that the ability of the model to run on a platform with limited capabilities is more important than being accurate across different operating conditions, which informs decisions related to model design and training. 

%% file: application.tex
\section{Application in Practice}\label{sec:application}

The proposed approach for ML model test case generation has been integrated into MLTE (ML Test and Evaluation) — a system-centric, QA-driven, semi-automated process and infrastructure that enables negotiation, specification, and testing of ML model and system qualities \cite{Maffey2023}\cite{MLTE2023}. 

MLTE has been adopted by several modeling teams in an organization that develops mostly computer vision and prediction models for a variety of field-operated systems. The motivation for MLTE adoption was that many of the organization's models were failing during integration, testing, and operations, which they attributed to the lack of a systematic model test and evaluation (T\&E) process. In addition, they often work with very vague system requirements and have to make assumptions about the system and operational environment characteristics, which contributed to failures. 

Feedback from model developers and product owners on MLTE and its QA-driven requirements elicitation and test case generation process indicates that MLTE:
\begin{itemize}
    \item Forces model developers trained as data scientists to think beyond model performance 
    \item Identifies a larger number of relevant model and systems requirements
    \item Facilitates the development and reuse of test code for these requirements.
    \item Ensures that failures and inconsistencies in model performance are captured during testing, rather than finding issues in production
    \item Provides evidence of testing which increases trustworthiness of ML models
\end{itemize}

%% file: related-work.tex
\section{Related Work}\label{sec:related-work}

With respect to ML model testing, most research focuses almost exclusively on model performance \cite{Braiek2020}, neglecting whether the model aligns with system goals and is ready for use in production \cite{Golendukhina2022}. There is substantial research on improving model testing methods (\eg \cite{Ribeiro2020}\cite{Yang2022}) and evaluating model properties such as robustness \cite{Goodfellow2018}, fairness \cite{Holstein2019}, security \cite{Boenisch2021}, safety \cite{Rismani2023}, trust \cite{Hopkins2021}, and explainability \cite{Bhatt2020}. Some recent studies propose requirements modeling connecting system and model requirements \cite{Nalchigar2021}\cite{Siebert2020}\cite{Villamizar2023}, which can be a foundation for testing. However, testing the model in the context of the system has largely been neglected in research. 

For test case generation, there is a large amount of research in generating test cases from requirements specifications and use cases dating back to late 1990s and early 2000s (\eg \cite{Offutt1999}), with some work focusing on automated generation (\eg \cite{Anand2013}), and as of late even using large language models (LLMs) (\eg \cite{Takerngsaksiri2024}). In practice, in addition to requirements and use cases, QA scenarios are often used to build test cases \cite{Bass2021}. However, to the best of our knowledge, using QA scenarios for test case generation for ML models is not an existing practice. 

The QA-driven approach described in this paper to specify ML model requirements and drive test cases from those scenarios, and integrated into a tool like MLTE, fills all these gaps by connecting system requirements to model requirements to test cases. 

%% file: summary.tex
\section{Summary and Next Steps}\label{sec:summary}

We presented an approach based on QA scenarios for the generation of test cases for ML models. This approach has been adopted in practice and integrated into MLTE --- a tool for system- and model-centric testing of ML models --- with very positive feedback on the benefits of the approach. Next steps for this work include (1) extending a MLTE artifact called the \textit{Negotiation Card} to more formally record QA scenarios, and (2) extending the \textit{Test Catalog} in MLTE with examples for additional quality attributes to further illustrate the approach, especially for model developers not trained in software architecture and software engineering practices.